\documentclass[11pt,a4paper]{article}
\usepackage{amsmath,amsfonts,amssymb,amscd,graphicx}
\catcode`\@=11
\@addtoreset{equation}{section}

\catcode`\@=12

\newtheorem{Proposition}{Proposition}[section]

\newfont{\gotico}{eufm10 scaled\magstephalf}
\newfont{\qvd}{msam10 scaled\magstephalf}

\def\demo{\par\noindent{\sc Proof. }\begingroup}
\def\enddemo{\hskip1em \mbox{\qvd \char3}\endgroup\par\medskip}

\def\iff{\Leftrightarrow}
\def\interior{\,\hbox{\vrule depth0pt height.6pt width4pt%
\vrule depth0pt height8pt}\;\,}

\def\de#1/de#2{\frac{\partial {#1}}{\partial {#2}}}
\def\De#1/de#2{\dfrac{\partial {#1}}{\partial {#2}}}

\def\det{{\rm det}\,}

\def\H{{\cal H}}
\def\E{{\cal E}}

\def\je{{\cal J}\/(\E)}

\def\he{\H\/({\cal C})}
\def\hc{\H\/({\cal C})}
\def\pc{\Pi\/({\cal C})}

\begin{document}
\vskip-2cm

\title{General Relativity as a constrained Gauge Theory}    
       
\author{Stefano Vignolo, Roberto Cianci\\
        DIPTEM Sez. Metodi e Modelli Matematici, Universit\`a di Genova \\
        Piazzale Kennedy, Pad. D - 16129 Genova (Italia)\\ 
        E-mail: vignolo@diptem.unige.it, cianci@diptem.unige.it
\and  
       Danilo Bruno\\
       Dipartimento di Matematica, Universit\`a di Genova \\
                Via Dodecaneso 35 - 16146 Genova (Italia) \\
                E-mail: bruno@dima.unige.it
}

\date{}              
\maketitle

\begin{abstract}{
The formulation of General Relativity presented in \cite{CVB3} and
the Hamiltonian formulation of Gauge theories described in \cite{CVB4} are made to interact.
The resulting scheme allows to see General Relativity as a constrained Gauge theory. 
}
\end{abstract}

\section{Geometrical preliminaries}
\noindent The recent developments in the study of ${\cal J}$-bundles \cite{CVB1,CVB2,VC}
allowed to build a first-order frame formulation of General Relativity
\cite{CVB3} on the one hand and a regular Hamiltonian formulation of gauge
theories \cite{CVB4} on the other. The interaction of the two aspects will allow to
deduce General Relativity as a constrained variational problem for a Hamiltonian function
of a $SO\/(1,3)\/$-gauge theory. 

\noindent For this purpose, the geometrical frameworks contained in \cite{CVB3} and
\cite{CVB4} will be
briefly revised: only the main results will be exposed, referring the reader to the cited
works for the details and the proofs.
 
\noindent First of all, the purely frame formulation of General relativity described in
\cite{CVB3} will
be considered. 

\noindent
Let $M$ be a $4$-dimensional space--time manifold,
allowing a metric $g$ with signature $(1,3)$. Let $P\to M$ be a principal fiber bundle
having structural group $G=SO(1,3)$ and $L(M)\to M$ be the co--frame bundle over $M\/$. 

\noindent
According to the gauge natural bundle framework (see \cite{FF} and references therein), the configuration space of the theory is the $GL\/(4,\Re)\/$-bundle $\pi:{\cal E}\to M$, associated
with $P\times_M L(M)$ through the left-action
\begin{equation}
\label{0.1}
\lambda : (SO(1,3)\times GL(4,\Re)) \times GL(4,\Re) \to GL(4,\Re), \qquad
\lambda\/(\Lambda,J;X) = \Lambda\cdot X \cdot J^{-1} 
\end{equation}
The space ${\cal E}$ can be referred to local coordinates $x^i,e^\mu_i$ ($i,\mu = 1,\ldots,4\/$), subject to the following transformation laws
\begin{equation}
\label{0.2}
\bar{x}^i =\bar{x}^i\/(x^j), \qquad \bar e^\mu_j = e^\sigma_i
\Lambda^\mu_{\;\;\sigma}(x)\de x^i/de{\bar x^j}
\end{equation}
with $\Lambda^\mu_{\;\;\sigma}(x) \in SO(1,3)\;\forall\,x\in M$. 

\noindent
The dynamical fields of the theory are sections $\gamma :x^i \to (x^i,e^\mu_j\/(x^i))\/$ of the bundle ${\cal E}\to M\/$. Every section $\gamma\/$ can be thought as a family of local sections of $L\/(M)\to M\/$, glued to each another by Lorentz transformations. Any such a section induce a corresponding metric on $M\/$, defined as $g_{ij}\/(x^k) := \eta_{\mu\nu}e^\mu_i\/(x^k)e^\nu_j\/(x^k)\/$, where $\eta_{\mu\nu}=\eta^{\mu\nu} :=diag\/(-1,1,1,1)\/$. 

\noindent Moreover, let ${\cal C}\to M$ denote the space of principal connections on $P$, 
consisting in the quotient bundle $j_1\/(P,M)/G$. 
A set of local coordinates over ${\cal C}$ is provided by the functions
$x^i,\omega_i^{\;\;\mu\nu}\,(\mu<\nu)$, subject to the following transformation
laws
\begin{equation}\label{1.1}
\bar{x}^i =\bar{x}^i\/(x^j), \qquad \bar{\omega}^{\;\;\mu\nu}_i =
\Lambda^\mu_{\;\;\sigma}\/(x)\Lambda^\nu_{\;\;\gamma}\/(x)\de x^j/de {\bar x^i}
\omega^{\;\;\sigma\gamma}_j - \Lambda_{\;\;\sigma}^\eta\/(x)\de \Lambda^\mu_{\;\;\eta}\/(x)/de {x^h}
\de x^h /de{\bar x^i} \eta^{\sigma\nu}
\end{equation}
where $\Lambda^\mu_{\;\;\nu}(x)\in SO(1,3)\; \forall\,x \in M$ and $\Lambda_{\sigma}^{\;\;\nu}:= \left(\Lambda^{-1}\right)^\nu_{\;\;\sigma} =\Lambda^{\alpha}_{\;\;\beta}\eta_{\alpha\sigma}\eta^{\beta\nu}\/$. 

\noindent The velocity space of the theory is provided by the first ${\cal J}$-bundle of
$\pi:{\cal E}\to M$. It is built similarly to an ordinary jet--bundle, but the first order
contact between sections is calculated with respect to the exterior covariant differential (compare with 
\cite{CVB3} for the details). As far as this paper is concerned it is only needed to know
that: \\
1. The bundle $\je$ has all the properties of standard jet-bundles (compare
with \cite{CVB2}), such as contact 1-forms, raising of sections and vector fields.  \\
2. The bundle $\je\/$ is diffeomorphic to the fiber product $\E\times_{M}{\cal C}\/$ over $M\/$. $\je\/$ can be then referred to local coordinates $x^i,e^\mu_i,\omega^{\;\;\mu\nu}_j$ as above. In such coordinates, a section $\gamma : M\to\je\/$, $\gamma : x^k \to (x^k,e^\mu_i\/(x^k),\omega_i^{\;\;\mu\nu}\/(x^k))\/$, is said holonomic --- or kinematically admissible --- if the quantities $\omega_i^{\;\;\mu\nu}\/(x^k)\/$ are the coefficients of the spin connection generated by the metric $g_{ij}\/(x^k)=\eta_{\mu\nu}e^\mu_i\/(x^k)e^\nu_j\/(x^k)\/$.\\
3. The variational problem built on $\je$ through the $4$-form
\begin{equation}
\label{0.3}
\Theta = \frac{1}{4} \epsilon^{qpij} \epsilon_{\mu\nu\lambda\sigma} e^\mu_q e^\nu_p \left( d\omega_i^{\;\;\lambda\sigma} \wedge ds_j + \omega_{j\;\;\;\eta}^{\;\;\lambda} \omega_i^{\;\;\eta\sigma} ds \right)
\end{equation}
with $ds := dx^1\wedge\ldots\wedge dx^4\/$ and $d\/s_j := \de /de {x^j} \interior ds$, provides the following field equations for critical sections $\gamma : x^k \to (x^k,e^\mu_i\/(x^k),\omega_i^{\;\;\mu\nu}\/(x^k))\/$
\begin{subequations}
\begin{equation}
\label{0.3a}
\epsilon^{qpij}\epsilon_{\mu\nu\lambda\sigma}e^\mu_q\left(\de e^\nu_p /de{x^j} + \omega^{\;\;\nu}_{j\;\;\;\rho} e^\rho_p \right) = 0 
\end{equation}
\begin{equation}
\label{0.3b}
\frac{1}{2} \epsilon^{qpij}\epsilon_{\mu\nu\lambda\sigma} e^\mu_q \left(\de \omega_i^{\;\;\lambda\sigma} /de {x^j} + \omega^{\;\;\lambda}_{j\;\;\;\eta}\omega^{\;\;\eta\sigma}_{i} \right) = 0 
\end{equation}
\end{subequations}
deriving from the Euler--Lagrange equations 
\[
\gamma^*\/(X\interior d\Theta) =0 \quad\quad \forall X\in D^1\/(\je)
\]
associated with the form $\Theta\/$ through usual vanishing boundary conditions.

\noindent Eqs.~\eqref{0.3a} ensure the kinematic admissibility of the critical sections,
expressed as 
\[
\de e^\nu_p /de{x^j}\/(x) - \de e^\nu_j/de{x^p}\/(x) = \omega^{\;\;\nu}_{p\;\;\;\rho}\/(x)e^\rho_j\/(x)  - \omega^{\;\;\nu}_{j\;\;\;\rho}\/(x)e^\rho_p\/(x)
\]
allowing us to identify the components $\omega^{\;\;\nu}_{p\;\;\;\rho}\/(x)$ with the
coefficients of the spin--connection associated with the metric $g_{ij}(x) =
\eta_{\mu\nu}e^\mu_i\/(x)e^\nu_j\/(x)$. 

\noindent Taking the previous result into account, eqs.~\eqref{0.3b} are equivalent to
Einstein equations (provided that $\det\/(e^\mu_i)\not =0\/$), written in the form
\[
\frac{1}{4} \epsilon^{qpij}\epsilon_{\mu\nu\lambda\sigma}e^\mu_p\/(x)
R_{ji}^{\;\;\;\lambda\sigma}\/(x) = 0 
\]
where 
\[
R_{ji}^{\;\;\;\lambda\sigma}\/(x) = \de \omega_i^{\;\;\lambda\sigma} /de {x^j}\/(x) - \de
\omega_j^{\;\;\lambda\sigma} /de {x^i}\/(x) +
\omega^{\;\;\lambda}_{j\;\;\;\eta}\/(x)\omega^{\;\;\eta\sigma}_{i}\/(x)  -
\omega^{\;\;\lambda}_{i\;\;\;\eta}\/(x)\omega^{\;\;\eta\sigma}_{j}\/(x)  
\]
denotes the curvature tensor of the metric $g$. 

\noindent As a result, the geometrical framework $(\je,\Theta)$ is the natural setting to
build a variational first--order purely frame (or, equivalently, a zero--order frame--affine) formulation of General Relativity.

\noindent
On the other hand, the Hamiltonian framework for gauge theories has been
described in \cite{CVB4}. Given a principal bundle $P\to M$, which takes the
gauge-invariance into account, it is possible to create a geometrical framework where
gauge theories can be described as {\it non-singular} Lagrangian and Hamiltonian theories. 
The argument consists in taking the space ${\cal C}\to M$ of connection $1$-forms over $P$
into account and building its first ${\cal J}$-bundle. This structure
is best described in \cite{CVB2}; in particular it is there shown how the components $R^{\;\;\;\mu\nu}_{ij}$ of the
curvature forms can be chosen as coordinates over the fibers of ${\cal J}\/({\cal C})$.
The use of the ${\cal J}$-bundle allows to build a {\it regular} Lagrangian theory. 

\noindent From now on, a review of the argument is proposed. Nevertheless, it will be
adapted to the current situation, where the gauge group is $SO(1,3)$.  

\noindent 1. A set of local coordinates on ${\cal J}\/({\cal C})$ is provided by the functions
$x^i,\omega_j^{\;\;\mu\nu},R_{ij}^{\;\;\;\mu\nu}\/$ $(\mu<\nu,\,i<j)\/$, subject to the transformation laws \eqref{1.1}
and
\[
\bar R_{ij}^{\;\;\;\mu\nu} = R_{rs}^{\;\;\;\lambda\sigma}\de x^r /de{\bar x^i} \de x^s /de{\bar x^j}
\Lambda^{\mu}_{\;\;\lambda} \Lambda^{\nu}_{\;\;\sigma}
\]

\noindent 2. The Hamiltonian theory is built considering the module
$\Lambda^4\/({\cal C})$ of $4$-forms over $\cal C$
and its sub--bundles $\Lambda^4_r\/({\cal C})$ consisting of those forms vanishing when $r$ of
their arguments are vertical vectors. The whole argument is described in a quite general
setup in \cite{CVB4}. 

\noindent In particular, the following submodules are taken into account:
\begin{equation}
\label{2.1}
\Lambda_1^4({\cal C}):=\{\alpha\in\Lambda^4({\cal C}) : \alpha = p(\alpha)\,ds\}
\end{equation}
and
\begin{equation}
\label{2.2}
\Lambda_2^4({\cal C}):=\{\alpha\in\Lambda^4({\cal C}) : \alpha = p(\alpha)\,ds + \frac{1}{2}\Pi_{\;\;\mu\nu}^{ji}(\alpha)\,d\/\omega^{\;\;\mu\nu}_{\,i}\wedge d\/s_j\}
\end{equation}
The sets of functions 
$\{x^i,\omega^{\;\;\mu\nu}_i,p\}$ and $\{x^i,\omega^{\;\;\mu\nu}_i,p,\Pi_{\;\;\;\mu\nu}^{ij} \}$ ($\mu<\nu\/$) form  systems of local
coordinates on $\Lambda_1^4\/({\cal C})$ and  $\Lambda_2^4\/({\cal C})$ respectively.

\noindent The bundle $\Lambda_2^4({\cal C})$ is endowed with the canonical Liouville $4$-form,
locally expressed as
\begin{equation}
\label{2.4} \Theta := p\,ds + \frac{1}{2}\Pi_{\;\;\;\mu\nu}^{ji}\,d\/\omega^{\;\;\mu\nu}_{\,i}\wedge d\/s_j
\end{equation}

\noindent
Being the bundle $\Lambda_1^4\/({\cal C})$ a vector sub-bundle of $\Lambda_2^4\/({\cal C})$,
the quotient bundle $\Lambda_2^4\/({\cal C})/\Lambda_1^4\/({\cal C})$ can be considered. The latter has
the nature of a vector bundle over ${\cal C}$ and is locally
described by the system of coordinates $x^i,\omega^{\;\;\mu\nu}_i,\Pi_{\;\;\;\mu\nu}^{ij}$, while the canonical projection makes $\pi: \Lambda_2^4\/({\cal C})\to
\Lambda_2^4\/({\cal C})/\Lambda_1^4\/({\cal C})$ into an affine bundle.

\noindent 3. The {\it phase space} of the theory is defined as the vector sub--bundle $\pc \subset
\Lambda_2^4\/({\cal C})/\Lambda_1^4\/({\cal C})$ consisting of those elements satisfying
\begin{equation}\label{2.4bis}
    \Pi_{\;\;\;\mu\nu}^{ij} (z) = - \Pi_{\;\;\;\mu\nu}^{ji} (z)
\end{equation}
A local system of coordinates for $\pc$ is provided by
$x^i,\omega^{\;\;\mu\nu}_i,\Pi_{\;\;\;\mu\nu}^{ij}\/$ $(i<j,\mu<\nu)\/$, subject to the transformation laws (\ref{1.1}) together with (see \cite{CVB4})
\begin{equation}\label{2.3}
\bar \Pi_{\;\;\;\lambda\sigma}^{pq} = det\left(\de x^h /de{\bar x^k}\right) \Pi_{\;\;\;\mu\nu}^{ij} \Lambda^{\;\;\mu}_\lambda \Lambda^{\;\;\nu}_\sigma \de \bar x^q/de{{x}^j}\de \bar
x^p/de{{x}^i} 
\end{equation}
Besides, being $\pc$ a vector sub--bundle, the immersion $i :\pc \to
\Lambda_2^4\/({\cal C})/\Lambda_1^4\/({\cal C})$ is well defined and is locally represented by
eq.~\eqref{2.4bis} itself.

\noindent 4. The pull--back bundle $\hat \pi:\he\to\pc$ defined by the following commutative
diagram
\begin{equation}\label{2.4tris}
\begin{CD}
\he     @>{\hat i}>>    \Lambda_2^4\/({\cal C})    \\
@V{\hat \pi}VV              @VV{\pi}V  \\
\pc        @>i>>     \Lambda_2^4\/({\cal C})/\Lambda_1^4\/({\cal C})
\end{CD}
\end{equation}
will now be taken into account.  
The latter has the nature of an affine bundle over the
phase space. Every section $h:\pc \to \he$ is called a {\it
Hamiltonian section}, and is locally described in the form:
\begin{equation}\label{2.6} 
h: p = - H\/(x^i,\omega^{\;\;\mu\nu}_i,\Pi_{\;\;\;\mu\nu}^{ij}) - \frac{1}{16}\Pi_{\;\;\;\mu\nu}^{ij}\omega^{\;\;\lambda\sigma}_i \omega^{\;\;\rho\beta}_j
C^{\mu\nu}_{\rho\beta\,\lambda\sigma}
\end{equation}
where $C^{\mu\nu}_{\rho\beta\,\lambda\sigma}$ are the structure coefficients of the group $SO\/(1,3)\/$. In accordance with the literature, the function $H\/(x^i,\omega^{\;\;\mu\nu}_i,\Pi_{\;\;\;\mu\nu}^{ij})\/$ is called the {\it Hamiltonian\/} of the system.

\noindent The presence of the immersion $\hat i:\he\to\Lambda_2^4\/({\cal C})$, endows $\he$
with the canonical $4$-form $\hat i^*\/(\Theta)$, locally expressed as in
eq.~\eqref{2.4}. The latter will be simply denoted as $\Theta$ and will be called the
Liouville form on $\he$.

\noindent 5. The assignment of the Hamiltonian section allows to perform the pull-back of
the Liouville form on $\he$ to the phase space $\pc$. The result is a Hamiltonian
dependent $4$-form
\begin{equation}\label{2.7}
\Theta_h := h^*\/(\Theta) = -H(x^i,\omega^{\;\;\mu\nu}_i,\Pi_{\;\;\;\mu\nu}^{ij})\,ds - \frac{1}{2}\Pi_{\;\;\;\mu\nu}^{ij}
\/\left( d\/\omega^{\;\;\mu\nu}_{\,i}\wedge d\/s_j + \frac{1}{8}\omega^{\;\;\lambda\sigma}_i \omega^{\;\;\rho\beta}_j
C^{\mu\nu}_{\rho\beta\,\lambda\sigma}\,ds \right)
\end{equation}
Starting from the coordinate transformations \eqref{1.1} and \eqref{2.3}, it is easily seen that the form \eqref{2.7} is covariant. The reader is referred to \cite{CVB4} for a more general explanation.\\ 
The free variational problem performed on $\pc\/$ through the form \eqref{2.7}
provides the Hamilton--De Donder equations
\[
\gamma^*\/(X\interior d\Theta_h)=0 \quad\quad \forall X\in D^1\/(\pc)
\]
yielding final field equations for critical sections $\gamma : x^k \to (x^k,\omega_i^{\;\;\mu\nu}\/(x^k),\Pi^{ij}_{\;\;\;\mu\nu}\/(x^k))\/$ of the form 
\begin{subequations}\label{h1}
\begin{equation}
- \de H/de{\Pi_{\;\;\;\alpha\beta}^{ij}} - \de{\omega_i^{\;\;\alpha\beta}}/de{x^j} + \de{\omega_j^{\;\;\alpha\beta}}/de{x^i} - \frac{1}{4}\omega^{\;\;\nu\mu}_i \omega^{\;\;\rho\lambda}_j C^{\alpha\beta}_{\rho\lambda\, \nu\mu} = 0
\end{equation}
\begin{equation}
-\de H /de {\omega^{\;\;\mu\nu}_{i}} - \de \Pi_{\;\;\;\mu\nu}^{ji} /de {x^j} + \frac{1}{4}\Pi_{\;\;\;\lambda\sigma}^{ji} \omega^{\gamma\alpha}_j
C^{\lambda\sigma}_{\gamma\alpha\,\mu\nu} = 0
\end{equation}
\end{subequations}
\noindent 6. The acquired regularity of the Hamiltonian allows to build a non--singular inverse
Legendre transformation $Leg^{-1} : \pc \to {\cal J}\/({\cal C})\/$, described as
\begin{equation}\label{h2}
R^{\;\;\;\alpha\beta}_{st} = \de H/de{\Pi^{st}_{\;\;\;\alpha\beta}}
\end{equation}
The Lagrangian associated with the Hamiltonian $H\/$ can be obtained as
\begin{equation}\label{h3}
L\/(x^i,\omega_i^{\;\;\mu\nu},R_{ij}^{\;\;\;\lambda\sigma}) = \frac{1}{4}\Pi_{\;\;\;\lambda\sigma}^{ij}
R^{\;\;\;\lambda\sigma}_{ij} - H\/(x^i,\omega_i^{\;\;\mu\nu},\Pi^{ij}_{\;\;\;\lambda\sigma}\/(x^i,\omega_i^{\;\;\mu\nu},R_{ij}^{\;\;\;\mu\nu}))
\end{equation}
The latter gives rise to the Legendre transformation $leg : {\cal J}\/({\cal C})\to\pc\/$, expressed as
\begin{equation}\label{h4}
\Pi^{st}_{\;\;\;\alpha\beta} = \de L/de{R^{\;\;\;\alpha\beta}_{st}}
\end{equation}
Taking eqs. (\ref{h2}), (\ref{h3}) and (\ref{h4}) into account, the Lagrangian counterpart of the field equations (\ref{h1}) can be expressed as
\begin{subequations}\label{h5}
\begin{equation}
R^{\;\;\;\alpha\beta}_{ij} = - \de{\omega_i^{\;\;\alpha\beta}}/de{x^j} + \de{\omega_j^{\;\;\alpha\beta}}/de{x^i} - \frac{1}{4}\omega^{\;\;\nu\mu}_i \omega^{\;\;\rho\lambda}_j C^{\alpha\beta}_{\rho\lambda\, \nu\mu} 
\end{equation}
\begin{equation}
\de L /de {\omega^{\;\;\mu\nu}_{i}} - D_{j}\de L/de{R_{ji}^{\;\;\;\mu\nu}}=0
\end{equation}
\end{subequations}
$D_j\/$ denoting covariant differentiation. In particular, eqs. (\ref{h5}a) ensure the kinematic admissibility of the critical sections, so that eqs. (\ref{h5}b) represent actual Lagrange equations in gauge theory.

\section{Gauge Theory and General Relativity}
The actual object of the paper is to study the theory arising from the assignment of the following
Hamiltonian function 
\begin{equation}\label{1.3}
H = \Pi^{ij}_{\;\;\;\mu\nu}\Pi^{pq}_{\;\;\;\lambda\sigma}\eta^{\mu\lambda}\eta^{\nu\sigma}\epsilon_{ijpq}
\end{equation}
on the phase space $\Pi\/({\cal C})$. $\epsilon_{ijpq}\/$ denotes the Levi--Civita permutation symbol.

\noindent The Hamiltonian (\ref{1.3}) induces the corresponding Hamiltonian $4$-form (\ref{2.7}). Taking the explicit expression of the structure coefficients $C^{\mu\nu}_{\rho\beta\,\lambda\sigma}$ for the
group $SO(1,3)$ into account, we have
\begin{equation}\label{1.4}
\Theta_h = -H ds - \frac{1}{2}\Pi^{ij}_{\;\;\;\mu\nu}\left(d\omega_i^{\;\;\mu\nu}\wedge
ds_j + \omega_{j\;\;\;\lambda}^{\;\;\mu}\omega_i^{\;\;\lambda\nu} ds \right)
\end{equation}

\noindent The properties of the Hamiltonian \eqref{1.3} are described by the following
\begin{Proposition}\label{Prop1.1}
The Hamiltonian \eqref{1.3} is regular.
\end{Proposition}
\demo
One could prove the above statement through a direct calculation. More simply, it is easily seen that
the inverse Legendre transformation, induced by Hamiltonian \eqref{1.3},
\begin{equation}\label{1.5}
R^{\;\;\;\alpha\beta}_{st} = \de H/de{\Pi^{st}_{\;\;\;\alpha\beta}} = 8
\Pi^{pq}_{\;\;\;\lambda\sigma}\eta^{\alpha\lambda}\eta^{\beta\sigma}\epsilon_{stpq}
\end{equation}
is bijective, being its inverse function (the Legendre transformation) provided by
\begin{equation}\label{1.6}
\Pi^{ij}_{\;\;\;\lambda\sigma} = \frac{1}{32}R^{\;\;\;\alpha\beta}_{st}
\eta_{\alpha\lambda}\eta_{\beta\sigma}\epsilon^{stij}
\end{equation}
\enddemo

\noindent A comparison with eqs.~\eqref{h3}, \eqref{1.5} and \eqref{1.6} allows to
obtain the expression for the Lagrangian $L$ associated with the Hamiltonian
\eqref{1.3}: 
\[
L = \frac{1}{256} R^{\;\;\;\alpha\beta}_{st} R^{\;\;\;\lambda\sigma}_{ij}
\eta_{\alpha\lambda}\eta_{\beta\sigma}\epsilon^{stij}
\]

\noindent Now, let us define a map $i:\je\to \Pi\/({\cal C})$, locally expressed as
\begin{equation}\label{1.8}
\left\{
\begin{aligned}
x^i &= x^i\\
\omega_i^{\;\;\mu\nu} &= \omega_i^{\;\;\mu\nu} \\
\Pi^{ij}_{\;\;\;\lambda\sigma} &= - \frac{1}{2} e^\mu_q e^\nu_p \epsilon^{qpij} \epsilon_{\mu\nu\lambda\sigma} 
\end{aligned}
\right.
\end{equation}
Note that, according to the transformation laws \eqref{0.2} and \eqref{2.3}, the map \eqref{1.8} is well defined.\\ 
We have the following result
\begin{Proposition}
The map $i:\je\to \hc$, defined by eq.~\eqref{1.8}, is an immersion. 
\end{Proposition}
\demo
A direct calculation shows that
\[
\de \Pi^{ij}_{\;\;\;\lambda\sigma}/de{e^\alpha_k} = - e^\mu_p\epsilon^{kpij}\epsilon_{\alpha\mu\lambda\sigma}
\]
The kernel of the differential of the map \eqref{1.8} is determined by the
condition
\[
\de \Pi^{ij}_{\;\;\;\lambda\sigma}/de{e^\alpha_k} V^\alpha_k = 0 \quad \iff \quad e^\mu_p\epsilon^{kpij}\epsilon_{\alpha\mu\lambda\sigma} V^\alpha_k = 0 
\]
to be solved in the unknowns $V^\alpha_k$. 

\noindent A saturation with $e^\lambda_i$ and $e^\rho_j$ yields 
\[
e^\lambda_i e^\rho_j e^\mu_p\epsilon^{kpij}\epsilon_{\alpha\mu\lambda\sigma} V^\alpha_h e^h_\beta e^\beta_k   = 0 \quad \iff \quad e\; \epsilon^{\beta\mu\lambda\rho}\epsilon_{\alpha\mu\lambda\sigma} V^\alpha_h e^h_\beta = 0 
\]
which is equivalent to:
\[
\delta^\beta_\alpha \delta^\rho_\sigma V^\alpha_h e^h_\beta - \delta^\rho_\alpha \delta^\beta_\sigma V^\alpha_h e^h_\beta = 0 
\]
The above expression can be split into two cases:\\
i) $\;\rho\neq\sigma \quad \Rightarrow \quad V^\rho_h e^h_\sigma = 0 $ \\
ii) $\rho=\sigma \quad \Rightarrow \quad V^\alpha_h e^h_\alpha - V^\sigma_h
e^h_\sigma = 0 $ (the index $\sigma$ is not summed). 
Summing these four equations ($\sigma=1,\ldots,4\/$) one gets $3 V^\alpha_h
e^h_\alpha = 0$ (summed over $\alpha$) and therefore also $V_h^\sigma
e_\sigma^h = 0 \quad \forall\;\sigma=1,\ldots,4\/$ ($\sigma$ not summed).

\noindent We thus get that $V_h^\rho e_\sigma^h = 0 \quad \forall\;\sigma,\rho=1,\ldots,4\/$. A final saturation
with $e^\sigma_i$ yields $V_h^\nu = 0\;\forall\,h,\nu=1,\ldots,4\/$ as a result.
The application $i$ is therefore an immersion.
\enddemo

\noindent As a consequence of Proposition 2.2, the space $\je$ has the nature of an
immersed submanifold \cite{S} of $\pc$, fibered over $\cal C\/$. A simple calculation shows that the pull-back
$\Theta = i^*\/(\Theta_h)$ of the form \eqref{1.4} by means of the map
$i:\je\to\pc$ is the very same $4$-form \eqref{0.3} leading 
the regular variational approach to General Relativity on $\je$. 

\noindent The argument is based on the following 
\begin{Proposition}
The pull-back of the Hamiltonian \eqref{1.3} through the map $i$ vanishes identically, i.e.
$i^*\/(H) = 0 $. 
\end{Proposition} 
\demo
\[
\begin{aligned}
i^*\/(H) =& \frac{1}{4} e^\xi_s e^\eta_t \epsilon_{\xi\eta\mu\nu} \epsilon^{stij} e^\alpha_h e^\beta_k \epsilon_{\alpha\beta\lambda\sigma} \epsilon^{hkpq} \eta^{\mu\lambda} \eta^{\nu\sigma} \epsilon_{ijpq} = \\
=& \frac{1}{2} e^\xi_s e^\eta_t e^\alpha_h e^\beta_k \epsilon^{stij} \epsilon_{\xi\eta\mu\nu} \epsilon_{\alpha\beta\lambda\sigma} \left(\delta^h_i \delta^k_j - \delta^k_i \delta^h_j \right) \eta^{\mu\lambda} \eta^{\nu\sigma}  = \\ 
=& e^\xi_s e^\eta_t e^\alpha_h e^\beta_k \epsilon^{sthk} \epsilon_{\xi\eta\mu\nu} \epsilon_{\alpha\beta\lambda\sigma} \eta^{\mu\lambda} \eta^{\nu\sigma}  =  e\;\epsilon^{\xi\eta\alpha\beta}\epsilon_{\alpha\beta\lambda\sigma}\epsilon_{\xi\eta\mu\nu}\eta^{\mu\lambda}\eta^{\nu\sigma} = \\
=& 2 e\; \left( \delta^\alpha_\mu \delta^\beta_\nu - \delta^\alpha_\nu \delta^\beta_\mu \right) \epsilon_{\alpha\beta\lambda\sigma} \eta^{\mu\lambda} \eta^{\nu\sigma} = 4 e\;\epsilon_{\mu\nu\lambda\sigma}\eta^{\mu\lambda} \eta^{\nu\sigma} = 0 
\end{aligned}
\]
\enddemo

\noindent Summing all up, we have proved that the formulation of General Relativity proposed in \cite{CVB3} is deducible from a constrained variational problem in a $SO\/(1,3)$-gauge theory,
described by the Hamiltonian \eqref{1.3} and the constraint \eqref{1.8}.

\noindent
In this respect, we notice that, although the map $i\/$ is not one--to--one with its image, the constrained variational problem is well defined; this is due to the fact that if $z_1,z_2 \in \je\/$ and $i\/(z_1)=i\/(z_2)\/$, then we have $i^*_{z_1}\/(\Theta_h)=i^*_{z_2}\/(\Theta_h)\/$.

\end{document}